\documentclass[11pt,aps,superscriptaddress,preprintnumbers,amsmath,amssymb,nofootinbib]{revtex4}
\usepackage{epsfig}  
\usepackage{graphicx,color}
\usepackage{hyperref}
\usepackage{bm}
\usepackage{fourier}
\usepackage{multirow}

\hypersetup{
    colorlinks=true,
    citecolor=red,
    linkcolor=blue,
    filecolor=green,      
    urlcolor=magenta,
}
\usepackage{color}
\usepackage{float}
\usepackage{amsfonts}
\usepackage{amsmath}
\usepackage{slashed}
\usepackage{soul}

 \usepackage[normalem]{ulem}

 \definecolor{darkgreen}{cmyk}{1,0,1,0.4}
 
\definecolor{ultramarine}{rgb}{0.5, 0.125, 0.376}

\def\beq{\begin{equation}}
\def\eeq{\end{equation}}
\def\barr{\begin{array}}
\def\earr{\end{array}}

\def\mfa {\mathfrak A}
\def\mff {\mathfrak F}
\def\mfv {\mathfrak V}

\def\kst {K^{*0}}
\def\kstbar {\overline{\kst}}
\def\bar {\overline}
\begin{document} 

\title{Polarization fractions in $\bm{B\to V_1 V_2}$: U-Spin constraints and new physics signatures}

\author{Debajyoti Choudhury}
\email{debajyoti.choudhury@gmail.com}
\affiliation{Department of Physics and Astrophysics, University of Delhi, Delhi 110007, India}

\author{Suman Kumbhakar}
\email{kumbhakar.suman@gmail.com}
\affiliation{Department of Physics, University of Calcutta, 92 Acharya Prafulla Chandra Road, Kolkata 700009, India}

\author{Anirban Kundu}
\email{akphy@caluniv.ac.in}
\affiliation{Department of Physics, University of Calcutta, 92 Acharya Prafulla Chandra Road, Kolkata 700009, India}

\author{Soumitra Nandi}
\email{soumitra.nandi@gmail.com}
\affiliation{Department of Physics, Indian Institute of Technology, Guwahati 781039, India}

\begin{abstract}
We investigate the decays of $B$ mesons, {\em i.e.}, $B_d$, $B_s$, $B^+$, and their antiparticles, to two light vector mesons ($B \to V_1V_2$). We use the SU(2) U-spin symmetry, which relates $\Delta S = 0$ and $\Delta S = 1$ decay amplitudes through the interchange $d \leftrightarrow s$ and is an approximate symmetry of the Standard Model (SM), to relate the helicity amplitudes of these decays. Treating all the helicity amplitudes for these decays, and hence the reduced matrix elements, as free parameters, we find an acceptable solution within the SM, although this is driven by the fact that the number of observables is smaller than what is needed for a meaningful fit. To reduce the number of free parameters, we then use some apparently reasonable and theoretically motivated approximations, like the dominance of factorisable contributions over the non-factorisable ones, and hence a distinct hierarchy between the helicity amplitudes. We find that once the assumption of hierarchy is imposed, there is no solution, both in the exact U-spin limit as well as when substantial U-spin breaking is allowed for. The tension is primarily driven by the longitudinal polarisation fractions in almost all $\Delta S = 1$ decays, which are significantly smaller than the corresponding theoretical predictions based on the SM and U-spin symmetry. This is particularly true for $B_s \to K^{*0} \overline{K^{*0}}$, for which the individual disagreement with U-spin based expectation is more than $7\sigma$. Within the SM framework, the only effective resolution would be to entirely disregard the hierarchy between the longitudinal and transverse helicity amplitudes in the heavy-quark limit, as dictated by naive factorisation, implying that there must be large nonfactorisable contributions to all these decay amplitudes. We also explore whether some new physics (NP) in the $b\to s$ sector that does not respect the hierarchy among the helicity amplitudes can reduce the tension for all the $\Delta S=1$ modes. While the answer is partially in the affirmative, we find that for simplistic new physics scenarios, the tension still exists and the fit remains poor enough, if the hierarchy exists among the SM amplitudes. Some possible scenarios for a complete solution of the puzzle are also suggested.

 \end{abstract}
 
\maketitle

\section{Introduction}      \label{sec:intro}
Over the last couple of decades, several cases of apparent tension
between the Standard Model (SM) predictions and the experimental data
on hadronic $B$-meson decays have surfaced, and some of them are yet
to be resolved. One of the longstanding examples is the so-called $B
\to \pi K$ puzzle (for details, see, {\em e.g.},
Refs.~\cite{Beaudry:2017gtw,Kundu:2021emt} and references therein).
More recently, tensions have also been reported in $B$ decays to two
pseudoscalar
mesons~\cite{Amhis:2022hpm,Bhattacharya:2022akr,Grossman:2024amc,Berthiaume:2023kmp,Davies:2024vmv,BurgosMarcos:2025xja,Shi:2025eyp,Bhattacharya:2025wcq},
which are related by the SU(3)$_F$ flavour symmetry. A common
conclusion of these studies is that the measured branching ratios (BR)
and CP asymmetries ($A_{CP}$) show significant inconsistencies with
their SM predictions.  These discrepancies could be indicative of
either large (and unexpected) SU(3)$_F$-breaking effects, or possible
short-distance new physics (NP) contributions. Similar tensions have
also been observed in the two-body hadronic decays of the $D$
mesons~\cite{Schacht:2022kuj,Bause:2022jes,Bolognani:2024zno,Sinha:2025cuo}.

In this paper, we will focus on a tension that has surfaced rather
recently in $B\to V_1V_2$, {\em i.e.}, $B$ decays to two vector
mesons~\cite{LHCb:2019bnl}, particularly in the measurement of the
longitudinal polarisation fractions (LPF) in $B_{s,d} \to \kst
\kstbar$ decays\footnote{Some authors, like Ref.\ \cite{Yu:2024vlz}, argue that nonfactorisable effects like 
large final-state rescattering are non-negligible, but their contribution is still way too small to explain 
the unexpectedly low LPF in $B_s\to\kst\kstbar$.}. 
The recent results from the LHCb Collaboration \cite{LHCb:2025ftm}, which we will talk about a lot later in this
paper, have only helped in strengthening the tension to an almost unacceptable value.
These two decays, with identical final states, are
related by the SU(2) U-spin symmetry, characterised by the
$d\leftrightarrow s$ interchange, and hence relate the $\Delta S = 0$
($b\to d$) and $\Delta S = 1$ ($b\to
s$) amplitudes (see Refs.~\cite{Aleksan:2023rkc,Aleksan:2024dlf}).  Further
detailed analysis of these two particular decay
modes~\cite{Biswas:2023pyw} as well as possible NP
explanations~\cite{Lizana:2023kei,Biswas:2024bhn} are also available
in the literature.
Here, though, we adopt a broader outlook and try to encompass a
multitude of final states (with two vector mesons) that, pairwise, are
related by U-spin.  The advantages of using U-spin over conventional
quark-diagrammatic approaches may be found in detail in
Ref.\ \cite{Soni:2006vi}.  For us, the rationale for using the U-spin
symmetry, of course, is to relate the $\Delta S = 0$ and $\Delta S =
1$ amplitudes, so that they need not be treated as independent
variables\footnote{One could have used the method of topological amplitudes. However, there would have
been more independent variables than what we use in our approach.}.

The vector mesons that we consider in the final state are $\rho$
and/or $K^*$, both charged and neutral.  The SM amplitudes are
dominated by either $b\to d$ or $b\to s$ penguins; tree amplitudes,
wherever they exist, are suppressed. Thus, this is a fertile ground to
look for NP effects. Factorisation of the hadronic currents leads to
the fact that, in the infinite $b$-quark mass limit, both the final
state vector mesons must be produced in the longitudinally polarised
state, {\em i.e.}, with spin projection zero.  On the inclusion of
finite-$m_b$ effects, the final state mesons can also be transversely
polarised, with amplitudes relatively suppressed by factors of the
order of $\Lambda_{\rm QCD}/m_b$, with the QCD scale being of the
order of a few hundred MeVs.  This hierarchy between the longitudinal
and the transverse polarisation amplitudes holds within the context of
the SM, and may change if one considers a different Lorentz structure
for the relevant operators \cite{Beneke:2006hg}. Nonfactorisable
effects too may affect the hierarchy, but there is no {\em a priori}
way to reliably ascertain their relative importance. However, that they
are suppressed compared to the factorisable part is more or less an
accepted assumption.

The data, as shown in Table \ref{measurements}, show some interesting
features. The LPFs, denoted by $f_L$, are consistently smaller than
the corresponding theoretical expectations for all the $\Delta S = 1$
modes, although the discrepancy is most severe for
$B_s\to\kst\kstbar$. The tension exists even for $B_s\to\phi\phi$, a
mode without a U-spin partner channel, for which $f_L = 0.379 \pm
0.008$~\cite{LHCb:2014vmq,LHCb:2023exl}\footnote{The only exception is
$B^+\to \rho^0 K^{*+}$, for which $f_L = 0.720\pm 0.029$
\cite{LHCb:2025zvw}, but this mode is also without a U-spin partner.}.
One may try to argue that these decays are dominated by penguins in
the SM, and so may not be as clean as the tree-dominated decays; in
particular, that the nonfactorisable contributions can be large.  The
counterargument would be the numbers for the $\Delta S=0$ ($b\to d$)
decays, which are also penguin dominated but for which the LPFs are
closer to the SM predictions, except for the $B_d\to\kst\kstbar$ channel\footnote{It is
hard to envisage a scenario where the final state interaction is significant
only for this final state, so we do not discuss such a possibility.}.

Our goal in this paper will be to identify, in a quantitative way, how
serious the tension is between the LPF measurements for the $\Delta
S=1$ channels, in particular for $B_{s,d}\to \kst\kstbar$, and the
corresponding SM expectations. As the number of observables is limited,
one needs to make certain reasonable assumptions to check the robustness of
the SM dynamics. We will use two such assumptions, namely,
\begin{enumerate}
 \item 
 U-spin is an approximate symmetry of the SM. The amount of U-spin
 breaking should be of the order of the ratio of the meson decay constants,
 $f_K/f_\pi$, which is approximately 1.19. Thus, We expect U-spin to be broken
 by about 20\%, or to be even more conservative, 25-30\% at the most.
 \item 
 The factorisable parts of the amplitudes dominate over the non-factorisable parts.
 U-spin amplitudes include both of them, but if factorisation dominates, one expects 
 a hierarchy among the magnitudes of the helicity amplitudes, with the longitudinal
 helicity amplitude dominating over the transverse ones. If non-factorisable contributions
 are small, one may reasonably expect this hierarchy to be reflected in the U-spin
 amplitudes too, expressed in the helicity basis. We will quantify later what we mean by this
 statement.
\end{enumerate}

In the first part of this paper, we will try to see how good these assumptions are, or
whether the tension can be explained within the framework of the SM,
by relaxing some of these underlying assumptions. Such a relaxation, although 
not favoured in models like QCD factorisation, may not be too drastic, as 
nonperturbative QCD dynamics can still hold
some surprises. While including all the relevant modes (such as $B\to
V_1 V_2$, with $V_{1,2} = \{\rho, K^*\}$) in our analysis, we focus
mainly on those decays that are related by the U-spin
symmetry, with a moderate breaking. 
Thus, our analysis includes a total of eight decay modes,
namely, four with $\Delta S = 0$ and four with $\Delta S = 1$.  Among
these, two decays, {\em viz.}, $B_s \to K^{*-} \rho^+$ and $B_s \to
K^{*-} K^{*+}$, have not yet been observed experimentally, but we will
predict the expected LPFs for them.  The decay amplitudes, in terms of
the U-spin reduced matrix elements (RME), are treated as free
parameters, and can be extracted from the available data. Note
  that in any other approach, like the one using topological diagrams,
  the number of free parameters is much larger, and the fit does not
  make much sense.

The observables that we use are, wherever available, CP-averaged branching ratios
(BR), CP asymmetries ($A_{CP}$), and polarisation fractions, mostly longitudinal, but a couple of
transverse fractions too. We find
that, in the exact U-spin symmetric limit, the fit is poor, with a
very high $\chi^2$, indicating a large tension with the SM. The main
source of the tension is the LPF of $B_s \to \kst\kstbar$,
which is only about one-fourth of that of its U-spin partner
channel, $B_d \to \kst\kstbar$.  Consequently, whereas U-spin symmetry is expected to be broken
in the SM at most at the level of 25-30\%, even a much larger
breaking is insufficient to significantly alleviate the tension. We find that the root cause of the tension
is the assumed hierarchy among the helicity amplitudes, if only the SM dynamics is at play. 

In the second part, we investigate whether this discrepancy could be a
signal of new physics (NP) beyond the SM.  The NP, of course, need not
respect U-spin. In fact, it can be very much flavour-specific. As an
instructive exercise, we perform a model-independent analysis,
introducing some four-fermion effective operators with various
possible Lorentz structures. As the tension is mostly confined to the
$\Delta S = 1$ sector, the NP operators that we introduce are those
that mediate $b\to s\bar{q}q$ decays, $q\in\{u,d\}$, leaving $\Delta S
= 0$ channels essentially uncontaminated by
NP, and hence explicitly breaking U-spin for NP\footnote{Note
  that while small WCs can be accommodated for $\Delta S = 0$ NP
  operators, these neither add anything to the global fit nor are
  necessary. For such flavour-specific NP, see, {\em e.g.},
Ref.\ \cite{Dighe:2007gt}.}.  What underlying theory leads to
  such NP structure is, however, a moot question; the only motivation
  for such a choice is the data --- this is entirely a data-driven
  analysis. One may note that only in SM and theories with minimal
  flavour violation, the $b\to d$ transition strength is suppressed
  compared to the $b\to s$ one, this need not be true for NP models in
  general.  For the SM
part of the amplitudes, we stick to the two assumptions enlisted
before.  While the domination of the LPF is expected for the SM part
of the $\Delta S=1$ amplitudes, the NP operators, particularly scalar
and tensor types, lead to a significant transverse polarisation
fraction. We show that this leads to a softening of the tension, but
not to a complete explanation. We also discuss qualitatively what
happens when one introduces NP in $b\to d$ transitions too.

The paper is organised as follows. In Section \ref{sec:u-spin-theory}, we briefly discuss how U-spin relates various
$B\to V_1V_2$ amplitudes in the SM. This is followed by an analysis within the framework of SM, where we show
quantitatively how large the tension is, and pinpoint its source. In Section \ref{sec:NP}, 
we introduce various NP operators, and again do the analysis, taking the NP only in the $\Delta S = 1$ sector.
However, both $\Delta S=0$ and $\Delta S=1$ amplitudes
related by U-spin are taken into account for the SM part of the decay channels. 
We find that several possible NP operators can 
help in partially resolving the tension. We summarise and conclude in Section \ref{sec:conclusion},
 and relegate a few necessary details to the Appendix.


\section{U-spin analysis}      \label{sec:u-spin-theory}
%
The U-spin group SU(2)$_U$ is a sub-group of the full SU(3)$_F$ symmetry, with $(d,\, s)$ and 
$(\bar{s},\, -\bar{d})$ in the fundamental representation,
and all others quark fields $(u, c, b, t)$ as singlets. Of the spin-1 SU(3)$_F$ octet given by
\beq
\barr{rcl c rcl c rcl c rcl}
K^{*+} & = & u\bar{s}, &\quad&
K^{*-} & = & \bar{u}s, &\quad&
\rho^+ &= & -u\bar{d}, &\quad &
\rho^- &= & \bar{u}d, \quad
\\[1ex]
\kst &= & d\bar{s}, & \quad &
\kstbar &= & -s\bar{d},&\quad &
U^{*0} & = & \frac{1}{\sqrt{2}}\left(s\bar{s}-d\bar{d}\right),
&\quad &
U^*_8 & = & \frac{1}{\sqrt{6}}\left(2u\bar{u}-d\bar{d}-s\bar{s}\right)\,,
\earr
\eeq
under SU(2)$_U$, $\left(K^{*+}, \, \rho^+\right)$ and $\left(\rho^-,\,
K^{*-}\right)$ are doublets, $\left(\kst,\,U^{*0},\, \kstbar\right)$
form a triplet, while $U^*_8$ is a singlet. The ninth vector meson
$U^*_1 = \frac{1}{\sqrt{6}}\left(u\bar{u}+d\bar{d}+s\bar{s}\right)$
is, of course, a singlet under the entire SU(3)$_F$. The states
$U^{*0}$, $U^*_8$ and $U^*_1$ are linear combinations of the vector
mesons $\rho^0$, $\phi$ and $\omega$.  Similarly, $B^+$ is a U-spin singlet while
$\left(B_d,~B_s\right)$ form a U-spin doublet.

The decays $B\to V_1V_2$, where $B \in \left\{B^+,B_d,B_s\right\}$ and
$V_1,V_2\in \left\{\rho, K^*\right\}$, including all charged and
neutral states, can be classified into two broad classes, namely,
strangeness-conserving $\Delta S=0$ $\left(\bar{b}\to
\bar{d}q\bar{q}\right)$ and strangeness-violating $\Delta S = 1$
$\left(\bar{b}\to \bar{s}q\bar{q}\right)$, with $q\in
\left\{u,d,s\right\}$. The effective weak Hamiltonian for these decays
can be written as a sum of local operators $\mathcal{O}_i$ multiplied
by short-distance Wilson coefficients (WC) $C_i$, and products of
relevant Cabibbo-Kobayashi-Maskawa (CKM) matrix elements, $\lambda^D_p
= V^*_{pb}V_{pD}$, where $p\in\left\{u,c,t\right\}$ and $D\in\{d,
s\}$, depending on the decay mode under consideration. $C_i$ and the
expectation values of $\mathcal{O}_i$ are functions of the
renormalisation scale $\mu$, for which we adopt the natural
choice\footnote{Varying $\mu$ in the window $(\textstyle\frac12 m_b, 2 m_b)$ brings in
only relatively minor quantitative changes in our conclusions.}
$\mu=m_b$.  Exploiting the unitarity of the CKM matrix, we can write
the Hamiltonian as\footnote{Note that the CKM factors, $\lambda^D_{u}
= V^*_{ub}V_{uD}$ and $\lambda^D_{c} = V^*_{cb}V_{cD}$, involve only
the $u$ and $c$ quarks, with the top-quark contribution having been
appropriately subsumed using the unitarity condition $\lambda^D_{u} +
\lambda^D_{c}+ \lambda^D_{t} = 0 $.}
\begin{eqnarray}
    \mathcal{H}^{\bar{b}\to \bar{D}}_{\rm eff} = \frac{G_F}{\sqrt{2}}\sum_{p=u,c} \lambda^{D}_{p} \left( C_1
    \mathcal{O}^p_{1D} + C_2\mathcal{O}^p_{2D} + \sum^{10}_{i=3}C_i\mathcal{O}_{iD}\right) + \mbox{h.c.} \,.
    \label{hamiltonian}
\end{eqnarray}
Here, $\mathcal{O}^p_{(1,2)D}$ are the left-handed current-current
operators (generated by tree-level $W$ boson exchange) that depend on
the identity of the up-type quark $p$, whereas $\mathcal{O}_{(3-6)D}$
and $\mathcal{O}_{(7-10)D}$ are the QCD and electroweak penguin
operators respectively, as defined in
ref~\cite{Beneke:2001ev}\footnote{We do not consider here the
electromagnetic and chromomagnetic penguin operators, commonly denoted
by ${\cal O}_{7\gamma}$ and ${\cal O}_{8g}$ respectively, as they are
not relevant for our discussion.}. From now on, we
will absorb the SM WCs into the definition of the operators, {\em
  i.e.}, define $C_i {\cal O}_i$ as just ${\cal O}_i$, so that the WCs
will not appear explicitly any further.

In the $\left\vert\, U \, U_3 \right\rangle $ basis, the
$\bar{b} \to \bar{d}$ Hamiltonian transforms like $\bar{d} \equiv
\left. -\vline \frac{1}{2}\, -\frac{1}{2} \right\rangle$ whereas the
$\bar{b} \to \bar{s}$ Hamiltonian transforms like $\bar{s}\equiv
\left. \vline \frac{1}{2}\, \frac{1}{2} \right\rangle$.  Using
the Wigner-Eckart theorem, it is straightforward to express the amplitudes for
  the $B\to V_1V_2$ decays in terms of the U-spin reduced matrix elements
  (RME)~\cite{Soni:2006vi}.  From now on, we will
write these amplitudes, derived using the Hamiltonian in
  Eq.~(\ref{hamiltonian}), in terms of the final state vector meson
helicity $h\in \{0,-,+\}$.

In terms of the U-spin RMEs defined as 
\begin{eqnarray}
    \mathcal{A}^{p,h}_{\frac{1}{2} D} = \left\langle \textstyle\frac12; h \,\vline\,\vline\,  \left[\mathcal{O}^{p}_{D}
    \right]^{\frac{1}{2}} \,\vline\,\vline\,0 \right\rangle , \quad 
    \mathcal{A}^{p,h}_{0 D} = \left\langle 0;h \,\vline\,\vline\,  
    \left[\mathcal{O}^{p}_{D}\right]^{\frac{1}{2}} \,\vline\,\vline\,\textstyle\frac12 \right\rangle, \quad 
    \mathcal{A}^{p,h}_{1 q} = \left\langle 1; h \,\vline\,\vline\,  \left[\mathcal{O}^{p}_{D}\right]^{\frac{1}{2}} 
    \,\vline\,\vline\,\textstyle\frac{1}{2} \right\rangle\,,
    \label{amp2}
\end{eqnarray}
the $\Delta S =0$ decay amplitudes, in units of $G_F/\sqrt{2}$, are
\begin{eqnarray}
    A_h(B^+\to K^{*+}\kstbar) & = & -\sqrt{\frac{2}{3}} \left[\lambda^u_{bd} \mathcal{A}^{u,h}_{\frac{1}{2} d}  + \lambda^c_{bd}  \mathcal{A}^{c,h}_{\frac{1}{2} d}\right]\nonumber\\
    A_h(B_d\to \kst\kstbar) &=& -\frac{1}{\sqrt{6}}  \left[\lambda^u_{bd}  \mathcal{A}^{u,h}_{0d}  + \lambda^c_{bd}  \mathcal{A}^{c,h}_{0d}\right]  -\frac{1}{2}  \left[\lambda^u_{bd} \mathcal{A}^{u,h}_{1d}  + \lambda^c_{bd}  \mathcal{A}^{c,h}_{1d}\right] \nonumber \\
    A_h(B_d \to \rho^+\rho^-) & =& \frac{1}{2} \left[\lambda^u_{bd}  \mathcal{A}^{u,h}_{0d}  + \lambda^c_{bd} \mathcal{A}^{c,h}_{0d}\right]  -\frac{1}{2}  \left[\lambda^u_{bd} \mathcal{A}^{u,h}_{1d}  + \lambda^c_{bd} \mathcal{A}^{c,h}_{1d}\right]\nonumber\\
    A_h(B_s\to \rho^+ K^{*-}) &=& \lambda^u_{bd} \mathcal{A}^{u,h}_{1d}  + \lambda^c_{bd} \mathcal{A}^{c,h}_{1d}\,,
\label{eq:ds0rme}
\end{eqnarray}
with\footnote{The top contribution has already been absorbed in the
definition of the operators.}
$p\in \{u,c\}$ and $D\in \{d,s\}$. The labels $0$ and
$\textstyle\frac12$ in the initial and final states denote their total
U-spin quantum numbers. With the initial state being spin-0, the helicities of the
daughters are equal and are denoted by a single label $h$.

Similarly, the $\Delta S =1$ amplitudes (again, in units of $G_F/\sqrt{2}$) are
\begin{eqnarray}
    A_h(B^+\to \kst \rho^{+}) &=& -\sqrt{\frac{2}{3}} \left[\lambda^u_{bs}  \mathcal{A}^{u,h}_{\frac{1}{2} s}  + \lambda^c_{bs}  \mathcal{A}^{c,h}_{\frac{1}{2} s}\right],\nonumber\\
    A_h(B_s\to \kst\kstbar) &=& -\frac{1}{\sqrt{6}}  \left[\lambda^u_{bs}  \mathcal{A}^{u,h}_{0s}  + \lambda^c_{bs} \mathcal{A}^{c,h}_{0s}\right]  -\frac{1}{2}  \left[\lambda^u_{bs} \mathcal{A}^{u,h}_{1s}  + \lambda^c_{bs} \mathcal{A}^{c,h}_{1s}\right] \nonumber \\
    A_h(B_s \to K^{*-}K^{*+}) & =& \frac{1}{2} \left[\lambda^u_{bs}  \mathcal{A}^{u,h}_{0s}  + \lambda^c_{bs} \mathcal{A}^{c,h}_{0s}\right]  -\frac{1}{2} \left[\lambda^u_{bs} \mathcal{A}^{u,h}_{1s}  + \lambda^c_{bs} \mathcal{A}^{c,h}_{1s}\right]\nonumber\\
    A_h(B_d\to K^{*+}\rho^-) &=&  \lambda^u_{bs} \mathcal{A}^{u,h}_{1s}  + \lambda^c_{bs} \mathcal{A}^{c,h}_{1s}\,.
    \label{amp1}
\end{eqnarray}
  
 However, one may note that not all the RMEs are independent. In the
 limit of exact U-spin \cite{Soni:2006vi}, the RMEs ${\cal A}^{p,h}_{nD}$,
 with $n\in\{0,\textstyle\frac12,1\}$, are related:
\begin{equation}
{\cal A}^{p,h}_{\frac12 d} = {\cal A}^{p,h}_{\frac12 s}\,,\ \ \ 
{\cal A}^{p,h}_{0 d} = {\cal A}^{p,h}_{0 s}\,,\ \ \  
{\cal A}^{p,h}_{1 d} =
{\cal A}^{p,h}_{1 s}\,, \ \ \ \forall~p,h\,,
\label{eq:u-relat}
\end{equation}  
rendering the $D$-label irrelevant.

The heavy quark symmetry, and the absence of large nonfactorisable
contributions, imply a certain hierarchy among the moduli of the
helicity amplitudes, which is in turn reflected in the hierarchy among the RMEs~\cite{Beneke:2006hg}:
\begin{equation}
  A_0 : A_- : A_+ \sim 1 : \frac{\Lambda_{\rm QCD}}{m_b} : \left(\frac{\Lambda_{\rm QCD}}{m_b}\right)^2\,.
  \label{hierarchy}
\end{equation}
It should be noted that Eq.\ (\ref{hierarchy}) is a joint consequence
of the $(V-A)$ structure of the weak interaction as well as the
assumption of naive factorisation being a good approximation.  As stressed before, 
we assume nonfactorisable contributions to the amplitudes
to be subdominant.  Under this assumption, the matrix elements of the
four-fermi operators in $\mathcal{H}_{\rm eff}$ are sought to be
approximated by products of two current matrix elements. In other
words, the helicity amplitudes $A_h^{V_1V_2}$ for the decay $B\to
V_1V_2$ are proportional to
\begin{eqnarray}
 A_h^{V_1V_2} \propto 
 \langle V^h_2 |(\bar{D}b)_{V-A}|\bar{B} \rangle \langle V^h_1|(\bar{q}q')_{V}|0 \rangle\,.
\end{eqnarray}
Using the form factors defined in Appendix~\ref{app}, we obtain
\begin{equation}
    A_0^{V_1V_2} \propto im^2_Bf_{V_1}\mfa_0^{B\to V_2}(0), \quad A_{\pm}^{V_1V_2} \propto im_B m_{V_1} f_{V_1}\mff_{\pm}^{B\to V_2}(0)
\end{equation}
where \begin{equation}
    \mff_{\pm}^{B\to V_2}\left(q^2\right) = \left(1+\frac{m_{V_2}}{m_B}\right)\mfa_1^{B\to V_2}\left(q^2\right) 
   \,  \mp \, \left(1-\frac{m_{V_2}}{m_B}\right)\mfv^{B\to V_2}\left(q^2\right)\,.
\end{equation}
The transverse amplitudes $A_{\pm}^{V_1V_2}$ are suppressed by a
factor of $m_{V_2}/m_B$ relative to the longitudinal amplitude
$A_0^{V_1V_2}$.  Moreover, in the heavy-quark limit, the vector and
axial-vector contributions to $\mff_+^{B\to V_2}(0)$ cancel due to an
exact form-factor relation~\cite{Beneke:2006hg}. This suppression
pattern leads to Eq.\ (\ref{hierarchy}).

These helicity amplitudes can also be expressed in terms of
transversity amplitudes such as 
\begin{equation}
    A_L = A_0, \quad A_{\parallel} = \frac{A_+ +A_-}{\sqrt{2}}, \quad A_{\perp} = \frac{A_+ -A_-}{\sqrt{2}}.
\end{equation}
In general, the amplitude for $B(p) \to V_1(q,\epsilon_1) V_2(k,\epsilon_2)$
decay can be written in the form~\cite{Valencia:1988it,Datta:2003mj}
\begin{eqnarray}
    \mathcal{M} = a\epsilon^*_1.\epsilon^*_2 + \frac{b}{m^2_{B}}(p.\epsilon^*_1)(p.\epsilon^*_2) + i\frac{c}{m^2_B}\varepsilon_{\mu\nu\alpha\beta}p^{\mu}q^{\nu}\epsilon^{\alpha}_1\epsilon^{\beta}_2\,,
\label{eq:bvv1}
\end{eqnarray}
where the form factors appear in the constants $a, b$, and $c$. 
The transversity amplitudes are related to the coefficients $a, b$, and $c$ by 
\begin{equation}
    A_{\parallel} = \sqrt{2}a, \quad A_{L} = -ax-\frac{m_1m_2}{m^2_B}b(x^2-1),\quad A_{\perp} = 2\sqrt{2}\frac{m_1m_2}{m^2_B}c\sqrt{x^2-1}\,,
 \label{eq:bvv2}
 \end{equation}
 where 
 \begin{equation}
 x = \frac{q.k}{m_1m_2} = \frac{m^2_B-m^2_1-m^2_2}{2m_1m_2}\,.
 \end{equation}
 Eq.\ (\ref{eq:bvv2}) shows why in the limit $m_B \gg m_1,m_2$, $A_L$
 dominates over $A_\perp$ or $A_\parallel$. The expressions are true
 if factorisation of the current-current operators hold, although
 even in the presence of nonfactorisable contributions, $A_L$
 still dominates.

In the naive factorisation limit, and using the values for the form
factors and masses, we find that for the modes under discussion,
$A_0:A_-:A_+ \simeq 1:0.27:0.01$ to a good approximation\footnote{One
may note that the hierarchy given in Eq.\ (\ref{hierarchy}) is much
stronger, even if we take $\Lambda_{\rm QCD}\sim 300$ MeV. However,
that is only an order-of-magnitude estimate.}, subject to the caveat
that this need not be valid~\cite{Beneke:2006hg} in the presence of
large nonfactorisable corrections.  Therefore, one should be able to safely ignore
$A_+$ for the SM analysis, but at the same time, it is worthwhile to
test this hierarchy experimentally.  Note that for operators with
other Lorentz structures, this hierarchy need not hold (even $A_+$ may
not be neglected), and this fact will be important in our discussion
of possible NP scenarios to explain the tension.

The number of independent RMEs, of the form ${\cal A}^{p,h}_{nD}$, with $p\in\{u,c\}$, $h\in\{0,-\}$, and
$n\in\{0,\textstyle\frac12,1\}$ are as follows:
\begin{itemize}
\item If all the helicity amplitudes are taken to be independent, and U-spin symmetry is not assumed, there are 
a total of 36 RMEs, 18 each for $\Delta S=1$ and $\Delta S = 0$ decays. 
\item If even an approximate U-spin symmetry is assumed, the $\Delta S=0$ and $\Delta S=1$ amplitudes get
related, and the number of independent amplitudes drops to 18. 
\item In the limit $A_+=0$, one has 24 independent RMEs without U-spin, and 12 in the limit of exact SU(2)$_U$. 
\end{itemize}

At this point, we would like to stress again that U-spin amplitudes
include both factorisable and nonfactorisable contributions, and are
not decomposed explicitly in terms of the form factors; they depend
only on the initial and final U-spin quantum numbers and not the
diagram topologies. The hierarchy among the amplitudes is not
inevitable in the U-spin approach. In the heavy quark limit, one
has $\mff_-/\mfa_0 \approx 1$ and $\mff_+/\mfa_0\approx \Lambda_{\rm QCD}/m_B$. These relations indicate
that the hierarchical structure of the amplitudes originates from the
relative sizes of the form factors contributing to them. Thus, any expectation of hierarchy among the U-spin
RMEs essentially means that the nonfactorisable contributions are assumed to be much smaller than the 
corresponding factorisable ones.

At the same time, U-spin is not an exact symmetry of nature, so one
should also check whether the polarisation puzzle is due to an
imperfect SU(2)$_U$ symmetry. One way to incorporate U-spin breaking
is to add an explicit breaking term in the $\Delta S = 1$
Hamiltonian. For the SM-based analysis that follows, we take a
different but equivalent route. We propose a possible mismatch of at
most 30\% --- as the breaking is expected to be moderate --- between
any $\Delta S=0$ RME and its $\Delta S = 1$ counterpart, as shown
later in Eq.\ (\ref{eq:ubreak1}). The mismatch need not be the same
for all U-spin related pairs. While we will show explicitly that any
such possible mismatch is not enough to address the polarisation
puzzle, the full analysis is detailed in Subsection
\ref{subsec:sm-analysis}.

Before moving further, let us note that one could have relaxed the assumption of $A_+=0$. 
In that case, there would have been 18 variables in the unbroken U-spin limit. With 18 observables, one could try for
an exact solution. We found that such an exact solution indeed exists---to be precise, very close to an exact 
solution, because all the RMEs are taken to be real, and $A_{CP}(B_d\to K^{*+} \rho^-)$ is not consistent
with zero at the $1\sigma$ level---but none of the 6 $A_+$s are negligibly small. In other words, the hierarchy does 
not exist, and an SM-only explanation desperately needs a large nonfactorisable contribution.

\subsection{Observables}    \label{subsec:observ}
\begin{table}[h]
\centering
\begin{tabular}{|l|l|}
\hline
$\Delta S = 0$ & $\Delta S = 1$\\
\hline
$B^+ \to \kstbar K^{*+}$ & $B^+ \to \kst \rho^+$ \\
$\mathcal{B}_{CP} = (0.91 \pm 0.29) \times 10^{-6}$ & $\mathcal{B}_{CP} = (9.2 \pm 1.5) \times 10^{-6}$  \\ 
& $A_{CP} = -0.01\pm 0.16$ \\
$f_L = 0.82^{+0.15}_{-0.21}$ & $f_L = 0.48\pm 0.08$ \\
\hline
$B_d\to \kst\kstbar $ & $B_s\to \kst\kstbar $\\
$\mathcal{B}_{CP} = (0.83\pm 0.24)\times 10^{-6}$ & $\mathcal{B}_{CP} = (11.1\pm 2.7)\times 10^{-6}$\\
$f_L = 0.600 \pm 0.028$ \cite{LHCb:2025ftm} & 
$f_L = 0.159\pm 0.012$ \cite{LHCb:2025ftm} \\
$f_\perp = 0.24\pm 0.03$ \cite{LHCb:2025ftm} 
& $f_\perp = 0.500 \pm 0.014$ \cite{LHCb:2025ftm} \\
\hline
$B_d\to \rho^+\rho^-$ & $B_s \to K^{*-}K^{*+}$ (not yet observed)\\
$\mathcal{B}_{CP} = (27.7\pm 1.9)\times 10^{-6}$ & $\spadesuit \, \mathcal{B}_{CP} = (1.27\pm 0.28)\times 10^{-5}$ \\
$A_{CP} = 0.00\pm 0.09$ & \\
$S_{CP} = -0.14\pm 0.13$ &\\
$f_L = 0.990^{+0.021}_{-0.019}$ & $\spadesuit \, f_L = 0.926\pm 0.063$ \\
\hline
$B_s \to K^{*-}\rho^+$ (not yet observed)& $B_d \to K^{*+}\rho^{-}$\\
$\spadesuit \, \mathcal{B}_{CP}= (2.96\pm 0.21) \times 10^{-5}$ &
$\mathcal{B}_{CP}= (10.3\pm 2.6)\times 10^{-6}$\\
& $A_{CP} = 0.21\pm 0.15$\\
$\spadesuit \, f_L = 0.972\pm 0.024$ & $f_L = 0.38\pm 0.13$\\
\hline
\end{tabular}
\caption{Experimental values of $B\to V_1V_2$ observables such as
  CP-averaged branching ratios ($\mathcal{B}_{CP}$), direct ($A_{CP}$)
  and indirect ($S_{CP}$) CP asymmetries, longitudinal ($f_L$) and
  transverse ($f_{\perp}$) polarisation fractions.  There are a total 18 measurements. These values are taken from
  Ref.~\cite{ParticleDataGroup:2024cfk}, unless indicated otherwise. The entries marked with
  $\spadesuit$ for the yet-to-be-observed modes are our predictions, made from the best fit values of the 
  RMEs (see text for details).}
\label{measurements}
\end{table}

To begin with, we list the observables, pertaining to various $\Delta
S=0$ and $\Delta S=1$ decays (into a vector boson pair) related by
SU(2)$_U$, that we would use in our analysis. These comprise (wherever
available) the CP-averaged branching fraction $\mathcal{B}_{CP}$, the
direct CP asymmetry $A_{CP}$, the mixing-induced CP asymmetry $S_{CP}$
and, most importantly, the polarisation fractions, $f_L$
(longitudinal), $f_\perp$ (transverse), and $f_\parallel$ (parallel),
and are defined as\footnote{Only two of the polarisation fractions are independent,
as their sum must be unity. Thus, we take only two of them as independent observables
(with correlation), even when data is available on all the three.}
\begin{eqnarray}
    \mathcal{B}_{CP} &=& 
    \frac{\sqrt{m^2_B-(m_{V_1}+m_{V_2})^2} \sqrt{m^2_B-(m_{V_1}-m_{V_2})^2}}{32\pi m^3_B\Gamma_B}
    \sum_{h=0,\pm}\left(|A_h|^2 + |\bar{A}_h|^2\right)\,,\nonumber\\
    A_{CP} &=& 
    \frac{\sum_{h=0,\pm}\left(|\bar{A}_h|^2 - |A_h|^2\right)}{\sum_{h=0,\pm}\left(|\bar{A}_h|^2 + |A_h|^2\right)}\,, 
    \nonumber\\
    S_{CP} &=& 
    2 {\rm Im}\left[e^{-2i\phi_M}\frac{\sum_{h=0,\pm} \bar{A_h}A^*_h}{\sum_{h=0,\pm}\left(|\bar{A}_h|^2 + |A_h|^2\right)}\right]\,,
    \nonumber\\
    f_{L,\perp,\parallel} &=& 
    \frac{1}{2} \left(\frac{|A_{L,\perp,\parallel}|^2}{|A_L|^2 + |A_{\perp}|^2 + |A_{\parallel}|^2} + \frac{|\bar{A}_{L,\perp,\parallel}|^2}{|\bar{A}_L|^2 + |\bar{A}_{\perp}|^2 + |\bar{A}_{\parallel}|^2}\right)\,.
\end{eqnarray}
Here $A_h$ and $\bar{A}_h$ are the helicity amplitudes for $B\to
V_1V_2$ and its CP conjugate decays, and $\phi_M$ is the weak phase of $B_d-\overline{B_d}$
mixing. The experimental data is summarised in Table~\ref{measurements}.
Among the 18 measurements, only one shows a striking tension with the SM, namely, $f_L (B_s\to\kst\kstbar)$, 
but $f_L$ for all $\Delta S=1$ modes are smaller than the corresponding theoretical expectations.  While
Eq.~(\ref{hierarchy}) predicts $f_L \gg f_\perp$, one finds, instead,
$f_L = 0.159\pm 0.012$ and $f_\perp = 0.500 \pm 0.014$ \cite{LHCb:2025ftm} for
$B_s\to\kst\kstbar$, in contradiction with Eq.~(\ref{hierarchy}). This
is also in contradiction with U-spin symmetry, as its U-spin partner
$B_d\to\kst\kstbar$ has $f_L = 0.600 \pm 0.028$. In the exact U-spin
limit, these two $f_L$s should be equal to each other, as predicted by
QCD factorisation~\cite{Aleksan:2023rkc}:
\begin{equation}
   \left[\frac{f_L(B_d\to \kst\kstbar)}{f_L(B_s\to \kst\kstbar)}\right]_{\rm QCDF} = 1.09^{+0.19}_{-0.08}\,,
\end{equation}
while we find 
\begin{equation}
    \left[\frac{f_L(B_d\to \kst\kstbar)}{f_L(B_s\to \kst\kstbar)}\right]_{\rm exp} = 3.77\pm 0.33\,, 
\end{equation}
corresponding to a tension of approximately $7.1\sigma$ between data and theory.  
This is the polarisation puzzle in the $B_{d,s}\to \kst\kstbar$ decays.

\subsection{Analysis within the SM}   \label{subsec:sm-analysis} 
We have made two assumptions so far: first, that SU(2)$_U$ is exact, and second, that the hierarchy of 
Eq.~(\ref{hierarchy}) is valid. Let us first check how far these assumptions are justified. First, let us assume
that the U-spin is an exact symmetry, but relax the hierarchy of
Eq.\ (\ref{hierarchy}). Thus, there are 18 independent RMEs, and as can be seen from Table \ref{measurements}, 
there are 18 observables. So, one expects an exact solution---at least almost an exact solution, as we have taken
all RMEs to be real, but $A_{CP}(B_d\to K^{*+}\rho^-)$ is not consistent with zero at the $1\sigma$ level. We perform a 
fit to the data, taking all the error bars at the $1\sigma$ level, and
that is precisely what we find, even when U-spin is taken to be unbroken, underlining that it is indeed a good 
symmetry of the flavour sector. We also predict the values of several observables, based on the fit, which are
shown in Table \ref{measurements}. 

This, however, does not mean that there is no tension with the SM, because the magnitudes of the helicity 
amplitudes do not even come close to respecting the hierarchy as shown in Eq.\ (\ref{hierarchy}). For 
example\footnote{The helicity amplitudes also include the CKM elements, and so even though these two decays 
are related by U-spin, the helicity amplitudes are not the same.}, 
\begin{eqnarray}
B_d\to \kst\kstbar &:& A_0 = (8.03\pm 1.65)\times 10^{-9}\,,\ \ A_- = (0.07\pm 1.80)\times 10^{-8}\,,\ \ A_+ = 
(0.67\pm 1.55)\times 10^{-8}\,,\nonumber\\
B_s\to \kst\kstbar &:& A_0 = (1.42\pm 0.37)\times 10^{-8}\,,\ \ A_- = (3.38\pm 7.54)\times 10^{-9}\,,\ \ A_+ = 
(3.30\pm 0.80)\times 10^{-8}\,.
\end{eqnarray} 

Next, we try to see how badly the hierarchy approximation is violated. For that, we still work in the exact U-spin limit
but take all $A_+=0$. This does not appear to be 
a drastic approximation, at least within the purview of the 
SM, as $A_+$ is anyway suppressed by $\left(\Lambda_{\rm QCD}\big/m_B\right)^2$.
This gives us altogether 12 RMEs, following Eq.\ (\ref{eq:u-relat}).
We assume these RMEs to be real, and perform a $\chi^2$ fit to the 18 measurements to extract
the best fit values of the 12 RMEs. For the minimisation, we use {\tt
  MINUIT} library~\cite{James:1975dr,James:1994vla}. The CP-violating
phases $\gamma$ and $\beta$ are taken as external constraints, with
current averages given in Table~\ref{input} along with the CKM matrix
elements.

 \begin{table}[h]
\begin{tabular}{||c|c|c||}
\hline\hline
$V_{ud} = 0.97367 \pm 0.00032$ & 
$V_{us} = 0.22431 \pm 0.00085$ & 
$V_{ub} = 0.00382 \pm 0.00020$ \\
$V_{cd} = 0.221 \pm 0.004$ &
$V_{cs} = 0.975 \pm 0.006$ &
$V_{cb} = 0.0411 \pm 0.0012$\\
\hline
$\gamma = \left(66.4^{+2.8}_{-3.0}\right)^{\circ}$ &
$\beta = \left(22.6^{+0.5}_{-0.4}\right)^{\circ}$ & \\
\hline\hline
\end{tabular}
\caption{The magnitudes of the CKM matrix elements \cite{ParticleDataGroup:2024cfk} and the CP violating 
phases~\cite{HFLAV:2022} used in the analysis.}
\label{input}
\end{table}
 
 We find a rather poor fit with $\chi^2/{\rm dof} = 37.8/6$ and the $p$-value of $1.23\times 10^{-6}$. 
 The fit being so poor, we do not even display the fit results. 
 However, one may immediately see that even this best
 fit does not respect the hierarchy between $A_0$ and $A_-$ 
 as shown in Eq.\ (\ref{hierarchy}). To demonstrate this, we have computed the ratios of the helicity amplitudes, 
 $|A_-/A_0|$,  using the best-fit values. For $\Delta S = 0$ decays, we find
\begin{equation}
   \left|\frac{A_-}{A_0}\right|_{B^+\to \kstbar K^{*+}} = 0.52 \pm 0.87\,, \quad
   \left|\frac{A_-}{A_0}\right|_{B_d\to \kst\kstbar} = 0.81\pm 0.64\,,\quad
   \left|\frac{A_-}{A_0}\right|_{B_d \to \rho^+\rho^-} = 0.10\pm 0.09\,.
\end{equation}
The hierarchy looks well-respected for the last one, and given the large error margins, the first two 
numbers are also consistent with the hierarchy.
The situation changes when we look at the $\Delta S = 1$ decays:
\begin{equation}
    \left|\frac{A_-}{A_0}\right|_{B^+\to \kst \rho^{+}} = 1.04\pm 0.19\,,\quad
    \left|\frac{A_-}{A_0}\right|_{B_s\to \kst\kstbar} = 2.39\pm 0.80\,,\quad
        \left|\frac{A_-}{A_0}\right|_{B_d\to K^{*+}\rho^-} = 1.30\pm 0.40\,.
\end{equation}

To ascertain that the poor fit is not due to the imposition of
an exact SU(2)$_U$, we insert a reasonable breaking: 
\begin{equation}
0.7 \leq \frac{{\cal A}^{p,h}_{nd}} {{\cal A}^{p,h}_{ns}} \leq 1.3 \ \ \ \forall p, h, n\,,
\label{eq:ubreak1}
\end{equation}
where we refer to Eqs.\ (\ref{amp2}) and (\ref{eq:u-relat}) for the
notation. This means a breaking of 30\% on either side of equality,
but the breaking need not be the same for different $p, h$, and
$n$. This feature has been incorporated in our analysis by randomising
the breaking within $\pm 30\%$ for different pair of U-spin related
RMEs. Thus, to take extreme values, while the ratio can be as low as
0.7 for a certain pair, it can be as high as 1.3 for another pair.
What we find is that the fit result remains almost equally poor; the
improvement is so marginal as to be almost imperceptible. 

We can even perform an analysis of $\Delta S=0$ and $\Delta S=1$
decays separately. This does not mean disregarding U-spin, as the
RMEs are still related by the symmetry, {\em e.g.}, in $B_d \to
\rho^+\rho^-$ and $B_s\to \rho^+ K^{*-}$. To ascertain whether the hierarchy is broken only for the $\Delta S=1$
modes, we impose $A_+=0$, $A_-=0.3 A_0$. For $\Delta S = 0$ decays, we have
6 parameters and 9 measurements, and get a rather poor fit with
$\chi^2/{\rm dof} = 144.1/3$, with $f_L(B_d\to \kst\kstbar)$ being the
most responsible one for the quality of the fit, contributing about 125 or 87\% to the total $\chi^2$.  
There is no minimum for the $\Delta S=1$ modes within the allowed ranges of the observables, as expected (which, in other words, means a disastrously poor $\chi^2$).

This just reaffirms what we have stated before: the tension is real
and rather severe if we stick to the hierarchy among the helicity
amplitudes. In other words, if there is no new physics beyond the SM, there must be
a large nonfactorisable contribution for $B_{s,d}\to\kst\kstbar$.
In the next Section, we will see whether NP can help.

\section{New Physics Interpretation}       \label{sec:NP}
Let us now entertain the possibility that new physics (NP) manifests itself in $b\to s$ decays alone, a choice while
not forced upon us, but is motivated by the fact that the tension is more severe for
$\Delta S=1$ decays. This simplifying assumption of a flavour-specific NP is, of course, not U-spin invariant. 
However, as we will soon see (and as could be gleaned from the preceding section), this lack of
U-spin symmetry is not the key ingredient. Rather, the resolution of the tension would depend on the introduction of 
specific $\Delta S = 1$ operators that may not respect the hierarchy of Eq.\ (\ref{hierarchy}).

The aforementioned U-spin violation could be motivated by
  postulating a heavy mediator $X$ with a tree-level (or effective)
  $\bar{b}sX$ vertex and also coupling to a light quark current. It
  needs to be borne in mind that such a flavour-violating coupling is
  constrained by $B_s$--$\bar{B_s}$ mixing. However, rather than
  committing ourselves to a specific ultraviolet completion, we
  consider, instead, to a  NP effective Hamiltonian at the
scale $m_b$, written as a sum of possible vector, scalar, and tensor operators:
\begin{equation}
    \mathcal{H}_{\rm NP} = \mathcal{H}_{\rm V} + 
    \mathcal{H}_{\rm S} + \mathcal{H}_{\rm T} + {\rm h.c.}\,,
\end{equation}
where
\begin{eqnarray}
{\rm Vector:} \quad \mathcal{H}_{\rm V} &=&
h_{v}e^{i\zeta_{v}}\left(\bar{q}_{\alpha}\gamma_{\mu}
(c_1+c_2\gamma_5)q_{\alpha}\right)\left(\bar{s}_{\beta}\gamma^{\mu}(c_3+c_4\gamma_5)b_{\beta}\right)\,,
\nonumber\\ {\rm Scalar:} \quad \mathcal{H}_{\rm S} &=&
h_{s}e^{i\zeta_{s}}\left(\bar{q}_{\alpha}(c_1+c_2\gamma_5)
q_{\alpha}\right)\left(\bar{s}_{\beta}(c_3+c_4\gamma_5)b_{\beta}\right)\,,
\nonumber\\ {\rm Tensor:} \quad \mathcal{H}_{\rm T} &=&
h_{t}e^{i\zeta_{t}}\left(\bar{q}_{\alpha}\sigma_{\mu\nu}
(c_1+c_2\gamma_5)q_{\alpha}\right)\left(\bar{s}_{\beta}\sigma^{\mu\nu}(c_3+c_4\gamma_5)b_{\beta}\right)\,.
 \label{eq:npdef1}
\end{eqnarray}
Here, $\zeta_{v,s,t}$ are the NP weak phases, $\alpha$, $\beta$ are colour indices, and $q\in \{u,d\}$.
The WCs $h_{v,s,t}$ encapsulate the NP coupling constants as well the suppression
$\Lambda^{-2}$, with $\Lambda$ being the scale at which the NP fields
have been integrated out (and constrained to be larger than the mass
of the heaviest such field). Nominally, we consider $\Lambda = 1$~TeV,
with the fit values of $h_{v,s,t}$ scaling as $(\Lambda/1~{\rm TeV})^2$. As for the coefficients $c_i$, they only determine
    the Lorentz structure of the current-current interactions, and we
    would consider only four possible combinations, {\em viz.}
   $(c_1,c_2,c_3,c_4) =
  (1,1,1,1), (1,-1,1,1), (1,1,1,-1)$ and $(1,-1,1,-1)$ for vector and
  scalar NP, and $(c_1,c_2,c_3,c_4) = (1,1,1,1)$ and $(1,-1,1,-1)$ for
  tensor NP.  Such a choice ensures that the relevant fermion fields are chiral.
  Furthermore, to keep the analysis tractable, we restrict ourselves to only one type of NP operator at a time, {\em
  viz.}  only a single set of the $c_i$s. Only with more data, would the  relaxation of these simplifying assumptions still be 
  amenable to a meaningful fit.

This is how we propose to do the analysis:
\begin{itemize}
 \item As already said, we will consider only one type of NP at a
   time.  Furthermore, we assume the corresponding $h_i$ to be real
   ({\em i.e.} the weak phase $\zeta_i = 0$), a choice motivated by
   the twin facts of the SM phase (arising almost entirely coming from
   $V_{tb}^* V_{ts}$) being very small and that there is hardly any
   tension in $A_{CP}$ for the $b\to s$ decays. In other words, the present data is completely consistent with
   the choice $\zeta_i=0$. This simplifying
   choice implies that, for a given combination of $c_i$ (a discrete
   set of choices), the number of free parameters has increased just
   by one.
   
   \item We will first concentrate on the three $\Delta S=1$ modes
     listed in Table~\ref{measurements}.  Since the SM amplitudes are
     supposed to respect U-spin, we would relate the SM contributions
     to the $\Delta S = 1$ amplitudes through an exact U-spin symmetry
     to those for the corresponding $\Delta S = 0$ modes which, by
     definition, do not feel NP.
 
    \item
 The $\Delta S = 1$ helicity amplitudes may be expressed as sum of the
 SM part and the NP part. For the SM part, we follow the hierarchy of Eq.\ (\ref{hierarchy}), assume
 $A_+=0$, and take $\left\vert A_-/A_0\right\vert = 0.3$.  For
   the NP part, we assume naive factorisation, and use
   Eqs.~(\ref{eq:bvv1}) and (\ref{eq:bvv2}) to obtain the NP parts of
   the helicity amplitudes $A_0$ and $A_\pm$ from the $a$, $b$, and
   $c$ coefficients which are functions of the NP coupling $h_i$. These are rather simplistic assumptions, 
   but our aim is to check whether the theoretically motivated SM dynamics, in the presence of NP, 
   can soften the tension. 

 \item
 As we have taken the NP to be confined to the $b\to s$ transition
 only, it is reasonable to redo the analysis with only the $\Delta
 S=1$ modes. The U-spin symmetry is no longer relevant, but we keep
 the hierarchy between $A_-$ and $A_0$ for the SM part of the
 amplitudes. Again, we find a very distinct improvement of the fit
 compared to the SM case\footnote{One has to compare with the SM fit
 where the hierarchy between $A_-$ and $A_0$ has been imposed.}.
 
 \item
 We also predict the LPF $f_L$ for the modes $B_s \to K^{*+}K^{*-}$
 and $B_s \to K^{*-}\rho^+$. For the latter, NP does not play a role,
 but the best fit values for the SM amplitudes change in the presence
 of NP and exact SU(2)$_U$.
\end{itemize}

Let us first calculate the NP contributions to the following $\Delta
S=1$ decays: $B^+\to \kst \rho^+$, $B_s\to \kst\kstbar$, and $B_d\to
K^{*+}\rho^-$. The quark-level processes are $b\to s\bar{d}d$ and
$b\to s\bar{u}u$. For the first channel, we get the following
expressions with Eqs.~(\ref{eq:bvv1}) and (\ref{eq:bvv2}) using naive
factorisation: 
\begin{itemize}
\item Scalar NP:
\begin{eqnarray}
    a &=& \frac{h_se^{i\zeta_s}}{4N_c} \left[(c_1c_4-c_2c_3)f_{K^{*0}}m_{K^{*0}} (m_{B} +m _{\rho})\mfa_1(q^2) 
    + c_1c_4f^T_{K^{*0}}\{(m^2_B-m^2_{\rho})g_{+}(q^2)+q^2g_{-}(q^2)\}\right] ,\nonumber\\
   \frac{b}{m_B^2} &=& -\frac{h_se^{i\zeta_s}}{4N_c}\left[(c_1c_4-c_2c_3)f_{K^{*0}}m_{K^{*0}} \frac{2\mfa_2(q^2)}{(m_{B} +m _{\rho})}
   +c_1c_4f^T_{K^{*0}}\{2g_{+}(q^2)+q^2g_0(q^2)\}\right],\nonumber\\
   \frac{c}{m_B^2} &=& \frac{h_se^{i\zeta_s}}{4N_c} \left[(c_1c_3-c_2c_4)f_{K^{*0}}m_{K^{*0}}\frac{2\mfv(q^2)}{(m_{B} +m _{\rho})})
   -2(c_2c_4+c_1c_3)f^T_{K^{*0}}g_{+}(q^2)\right]\,,
\end{eqnarray}

    \item Vector NP:
\begin{eqnarray}
    a &=& -\frac{h_ve^{i\zeta_v}}{2N_c} (c_1c_4+c_2c_3)f_{K^{*0}}m_{K^{*0}} (m_{B} +m _{\rho})\mfa_1(q^2)\,,\nonumber\\
   \frac{b}{m_B^2} &=& \frac{h_ve^{i\zeta_v}}{N_c} (c_1c_4+c_2c_3)f_{K^{*0}}m_{K^{*0}} \frac{\mfa_2(q^2)}{(m_{B} +m _{\rho})}\,,\nonumber\\
   \frac{c}{m_B^2} &=& \frac{h_ve^{i\zeta_v}}{N_c} (c_2c_4-c_1c_3) 
   f_{\kst} m_{\kst} \frac{\mfv(q^2)} {(m_{B} +m _{\rho}) }\,,
\end{eqnarray}

\item Tensor NP:
\begin{eqnarray}
    a &=& -\frac{h_te^{i\zeta_t}}{N_c} c_1c_4  f^T_{K^{*0}}\left[(m^2_B-m^2_{\rho})g_{+}(q^2)+q^2g_{-}(q^2)\right],\nonumber\\
   \frac{b}{m_B^2} &=& \frac{h_te^{i\zeta_t}}{N_c} c_1c_4 f^T_{K^{*0}}\left[2g_{+}(q^2)+q^2g_0(q^2)\right] ,\nonumber\\
   \frac{c}{m_B^2} &=& \frac{h_te^{i\zeta_t}}{N_c} 2c_1c_3  f^T_{K^{*0}}g_{+}(q^2)\,,
   \label{eq:tens}
\end{eqnarray}
\end{itemize}
where the $a$, $b$, and $c$ terms are contributions from NP only. 
Here $N_c$ is the number of colours, $f_{\kst}$ and $f^T_{\kst}$ are
the decay constants, and $\mfv$, $\mfa_1$, $\mfa_2$, $g_+$, $g_0$ and
$g_-$ are the $B\to \rho$ form factors, the definitions of which can
be found in Appendix~\ref{app}. For our analysis, we take $f_{K^*} =
0.204\pm 0.007$ GeV and $f^T_{K^*}/f_{K^*} = 0.712\pm 0.012$~\cite{RBC-UKQCD:2008mhs}.  
Similar expressions for other decay
channels may be obtained by suitable replacement of the daughter
mesons, and appropriate changes for the form factors, decay constants,
and masses. Our parametrisation of the NP operators, consistent with a
hypothetical flavour-changing charge-neutral colour-singlet mediator
that has been integrated out, leads to all the NP amplitudes being
colour-suppressed\footnote{We could have started with a NP
Hamiltonian that is a product of two colour-octet currents, and then
all our amplitudes would have been colour-allowed.}. 
 \begin{table}[h]
\begin{tabular}{||c|c|c|c||}
\hline
NP type  & Lorentz structure & best fit value ($h_i$) & $\chi^2/{\rm dof}$ \\
\hline
\multirow4*{Scalar}  & $(S-P)\otimes (S-P)$& $0.066\pm 0.005$ & 340.0/11 \\
& $(S-P)\otimes (S+P)$& $-0.168\pm 0.015$ & 198.4/11\\
  & $(S+P)\otimes (S-P)$& $0.132\pm 0.009$ & 484.2/11\\
  & $(S+P)\otimes (S+P)$& $0.068\pm 0.005$ & 400.9/11\\
  \hline
\multirow4*{Vector}  & $(V-A)\otimes (V-A)$& $0.051\pm 0.006$ & 622.6/11 \\
& $(V-A)\otimes (V+A)$& $0.107\pm 0.008$ & 565.7/11\\
  & $(V+A)\otimes (V-A)$& $-0.107\pm 0.008$ & 565.7/11\\
  & $(V+A)\otimes (V+A)$&  $-0.051\pm 0.006$ & 622.6/11 \\
\hline
\multirow2*{Tensor} & $(T \pm PT)\otimes (T-PT)$& $-0.029\pm 0.002$ & 163.6/11\\
  & $(T \pm PT)\otimes (T+PT)$ & $0.032\pm 0.002$ & 224.4/11\\
\hline
\end{tabular}
\caption{Fit results for different NP Lorentz structures taking NP only in $\Delta S =1$ decay assuming exact U-spin and $A_{-}/A_{0}=0.3$. The NP WCs $h_i$ are in TeV$^{-2}$ unit.}
\label{tabVI}
\end{table}



\begin{table}[h]
\begin{center}
\begin{tabular}{||c||c|c||c|c||}
\hline
\multirow2*{Mode} & \multicolumn{2}{c||}{$(S-P)\otimes (S+P)$} & 
\multicolumn{2}{c||}{$(T\pm PT)\otimes (T-PT)$}\\
\cline{2-5}
& 
$\left\vert A^{\rm NP}_-\right\vert / \left\vert A^{\rm SM}_-\right\vert$ &
$\left\vert A^{\rm NP}_0\right\vert / \left\vert A^{\rm SM}_0\right\vert$ & 
$\left\vert A^{\rm NP}_-\right\vert / \left\vert A^{\rm SM}_-\right\vert$ &
$\left\vert A^{\rm NP}_0\right\vert / \left\vert A^{\rm SM}_0\right\vert$
\\
\hline
\multirow3*{$B_s\to \kst\kstbar$} & \multicolumn{2}{c||}{$f_L = 0.18\pm 0.01$} & 
 \multicolumn{2}{c||}{$f_L = 0.18\pm 0.01$} \\
 & \multicolumn{2}{c||}{$f_\perp = 0.53\pm 0.01$} & 
 \multicolumn{2}{c||}{$f_\perp = 0.53\pm 0.01$} \\
 \cline{2-5}
& $1.08\pm 0.06$& $0.92\pm 0.05$ &  
$1.94\pm 0.63$ & $0.15\pm 0.05$\\
\hline
\multirow2*{$B^+\to \kst \rho^+$} & \multicolumn{2}{c||}{$f_L = 0.72\pm 0.04$} & 
\multicolumn{2}{c||}{$f_L = 0.42\pm 0.07$}
\\
\cline{2-5}
& $2.16\pm 0.30$ & $1.79\pm 0.25$ & 
$1.14\pm 0.15$ & $0.09\pm 0.01$\\
\hline
\multirow2*{$B_d\to K^{*+}\rho^-$} & \multicolumn{2}{c||}{$f_L = 0.78\pm 0.02$} & 
 \multicolumn{2}{c||}{$f_L = 0.39\pm 0.01$} \\
\cline{2-5}
& $2.31\pm 0.26$ & $1.94\pm 0.22$ &
$1.27\pm 0.33$ & $0.05\pm 0.01$ \\
\hline 
$\spadesuit$\ \ $B_s\to K^{*+}K^{*-}$
& \multicolumn{2}{c||}{$f_L = 0.90\pm 0.01$} & 
\multicolumn{2}{c||}{$f_L = 0.63\pm 0.05$}\\
\hline
\end{tabular}
\caption{The best-fit polarisation fractions for various $\Delta S=1$ modes and for some select Lorentz structures of the 
NP Hamiltonian. The two numbers shown immediately below $f_L$ (and $f_\perp$) are the NP and SM
transverse ($A_-$) and longitudinal ($A_0$) amplitude ratios. Note that $A_+$ (SM) has been taken to be zero
in the analysis, but $A_+$ (NP) is nonzero. We also show the prediction for $f_L\left(B_s\to K^{*+}K^{*-}\right)$, indicated by $\spadesuit$.}
\label{tabVIb}
\end{center}
\end{table}

When we take both $\Delta S=0$ and $\Delta S=1$ modes into account,
and impose (i) exact SU(2)$_U$ for the SM part of the amplitudes, and
(ii) the hierarchy $\left\vert A_-/A_0\right\vert = 0.3$ for all the
SM amplitudes, we have 18 measurements and only 7 free parameters: 6
SM RMEs plus one coming from the NP, and characterised by $h_i$, with
$i\in \{s,v,t\}$. The fit results are shown in Tables~\ref{tabVI} and
\ref{tabVIb}. The Lorentz structure of NP is shown as the product of
two currents; the first one is the $\bar{b}s$ current, and the second
one is the $\bar{q}q$ current. 

A look at the last column of Table~\ref{tabVI} immediately gives the impression that the fit, even with an 
order-of-magnitude improvement in $\chi^2$ from the corresponding SM fit just with the introduction of 
one single parameter, namely, the NP coupling $h_i$, still remains very poor. While this impression is not
wrong, we would like to analyse the individual contributions of different observables to the final $\chi^2$ 
to pinpoint the source of the tension. We have explicitly checked that a moderate breaking of SU(2)$_U$,
similar to what we have used before, hardly changes either the best fit values or the $\chi^2$. 

From Table~\ref{tabVI}, the most promising NP combinations are either scalar type: $(S-P)\otimes (S+P)$, 
with $\{c_1,c_2,c_3,c_4\}=\{1,-1,1,1\}$, or tensor type: $(T\pm PT)\otimes (T-PT)$, with $\{c_1,c_2,c_3,c_4\}=
\{1,\pm 1, 1,1\}$. For tensor operators, $c_2$ does not play role, as can be seen from Eq.\ (\ref{eq:tens}). Let us 
concentrate only on these combinations.

It is obvious that the more an operator can raise the transverse
polarisation fraction, the more successful it will be in bringing down
the $\chi^2$. However, note that the NP operators are only of $\Delta
S=1$ type. There is one observable in the $\Delta S=0$ channel---the
LPF of $B_d\to \kst\kstbar$---that contributes $\sim 125$ to $\chi^2$.
This was so for the SM fit, and this remains true for the SM$+$NP fit
too\footnote{Thus, one may reasonably argue that NP should also
  occur in $b\to d$ transitions. However, in a data-driven analysis
  like this, there is no way one can relate the strengths of $b\to s$
  and $b\to d$ transitions.}.  To highlight the improvement of the
fit for the $\Delta S=1$ modes, we perform the same exercise taking
only these channels into account. This naturally takes the U-spin out
of consideration. There are only 9 observables and 7 free parameters,
so we expect the fit to improve significantly. This is indeed the
case, and the total $\chi^2$ values for both the cases drop below 20.
We have also checked explicitly that introduction of a nonzero NP weak
phase $\zeta_i$ hardly changes the fit.  There is no significant
  tension with the CP-asymmetries for these modes, all of them being
  consistent with zero.  Even if we introduce a NP weak phase
  $\zeta_i$, the direct CP asymmetries will also depend on the strong
  phase difference of the relevant amplitudes, which
  are, {\em a priori}, unknown, and therefore this hardly puts any significant constraint on
  the NP weak phase $\zeta_i$. The mixing-induced CP asymmetry
  is expected to be tiny, as the $B_s$--$\overline{B_s}$ mixing
  amplitude is almost real within the SM. Thus, we do
  not lose anything by keeping all $\zeta_i=0$; it only simplifies the
  analysis.

Table \ref{tabVIb} also shows the ratios of the transverse ($A_-$) and
longitudinal ($A_0$) amplitudes for the NP and the SM. One may note
that the ratio is consistently larger than unity for the transverse
amplitudes and much smaller than unity for the longitudinal
amplitudes; in other words, there is no hierarchy between $A_-$ and
$A_0$ for the NP amplitudes. This leads to smaller $f_L$ values, in
agreement with the data.  Also note that $\left\vert A_0^{\rm
  NP}\right\vert \ll \left\vert A_0^{\rm SM}\right\vert$. With the BRs
being dominated by $A_0^{\rm SM}$, these are not significantly
affected by the NP and remain consistent with the experimental
numbers. 

 The Brs for the $B\to V_12V_2$ modes are used as inputs to the fit. 
 With the best fit values for the NP couplings $h_i$ --- for
  both $(S-P)\otimes (S+P)$ and $(T \pm PT)\otimes (T-PT)$ --- we have
  checked how far the BRs are affected even with final states that include
  pseudoscalar mesons. Our results are shown in Table \ref{tabBR}; all
  the modes, including those that are not shown explicitly, are
  consistent with the data within $1\sigma$, except
  $B_s\to\kst\kstbar$ which shows some tension. 

\begin{table}[h]
\begin{center}
\begin{tabular}{||c||c||c||c||}
\hline
Mode & Data ($\times 10^{6}$) & $(S-P)\otimes (S+P)$ ($\times 10^{6}$) & 
$(T \pm PT)\otimes (T-PT)$ ($\times 10^{6}$)\\
\hline
$B_s\to K^+K^-$ & $26.1\pm 1.6$ & $28.1\pm 3.2$ & $19.2\pm 3.8$ \\
$B_s\to K^0 {\overline{K}}^0$ & $17.6\pm 3.1$ & $25.2\pm 4.1$ & $17.0\pm 3.4$ \\
$B_s\to \kst\kstbar$ & $11.1\pm 2.7$ & $0.93\pm 0.06$ & $5.61\pm 0.93$\\
\hline
$B_d\to K^{*+}\rho^-$ & $10.3\pm 2.0$ & $9.93\pm 2.00$ & $9.55\pm 1.80$\\
$B_d\to \pi^- K^+$ & $20.0\pm 0.4$ & $18.5\pm 1.2$ & $18.8\pm 0.9$\\
$B_d\to \pi^0 K^0$ & $10.1\pm 0.4$ & $9.7\pm 0.7$ & $9.1\pm 0.9$\\
\hline
$B^+ \to \kst\rho^+$ & $9.2\pm 1.5$ & $7.32\pm 1.80$ & $10.34\pm 1.33$\\
$B^+\to \pi^+K^0$ & $23.9\pm 0.6$ & $20.5\pm 1.8$ & $21.2\pm 0.8$\\
$B^+\to \pi^0 K^+$ & $13.2\pm 0.4$ & $11.2\pm 0.9$ & $12.1\pm 0.5$\\
\hline
\end{tabular}
\caption{BRs, multiplied by $10^6$, for several modes affected by the NP operators discussed in the text.
The numbers for the experimental data are taken from Ref.\ \cite{ParticleDataGroup:2026aaa}.}
\label{tabBR}
\end{center}
\end{table}

One may improve the consistency marginally
with various measurements of $A_{CP}$ if $h_i$ is taken to be complex
(which in turn reduces $\chi^2$ further). This, however, is not needed
for the study of LPFs, and we have refrained from doing so.  It has
been checked that all other observables, {\em e.g.}, mass and width
differences in the $B_s$--$\overline{B_s}$ system and BR$(B\to
X_s\gamma)$, remain consistent with data even after NP effects are
included. 
  
Our assumption of the existence of NP only for the $\Delta S=1$ channels is rather simplistic. 
There is no symmetry that prevents such NP for the $\Delta S = 0$ channels, {\em i.e.}, for the $b\to d$
transitions. Even a basis rotation in the down-quark sector can induce such NP for $b\to d$. While we do 
not perform any detailed analysis of this scenario here, it is immediately obvious that this NP can reduce
the tension in $f_L(B_d\to\kst\kstbar)$ and may lead to an acceptable fit.

\section{Conclusions}          \label{sec:conclusion}
In this work, we perform a U-spin analysis of $B$ meson decays into
two light vector mesons $V_1V_2$, where $V_1,V_2\in \{\rho,
K^*\}$. These decays form a set of eight processes related by the $d
\leftrightarrow s$ exchange symmetry of U-spin. Out of these, we have
data on six modes. The available observables include CP-averaged
branching ratios, CP asymmetries, and polarisation fractions (mostly
longitudinal, but in a few cases, transverse too). 

Our motivation has been to probe the systematically low longitudinal
polarisation fractions (LPF) for $\Delta S=1$ decays, which is most
aggravated for the decay $B_s\to\kst\kstbar$. On the other hand, LPFs
for the $\Delta S=0$ decays are consistent with theoretical
expectation based on the SM dynamics, which predicts LPF to be close
to unity if nonfactorisable effects are neglected.

The rationale for using U-spin is as follows. Without U-spin, we would have lost the
  ability to express the amplitudes in a compact and structured
  form. In such cases, we would need to parametrise the various
  contributions based on topological classifications or adopt
  factorisation approaches, systematically accounting for
  non-factorisable corrections. In principle, non-factorisable
  contributions, such as annihilation or other subleading topologies,
  can also be interpreted in terms of U-spin symmetric or U-spin
  broken RMEs.  This interpretation reinforces the central role of
  U-spin symmetry, even when considering corrections and breaking
  effects.

To summarise the chain of reasoning, the decomposition of the
  amplitudes is carried out using U-spin symmetry, which forms the
  backbone of our analysis. While the U-spin symmetric approximation
  alone cannot resolve the polarisation puzzle, it provides the
  essential framework.  However, as U-spin is not an
  exact symmetry of nature, we have to incorporate the
  U-spin breaking effects. We found that even a sizable breaking (larger than the expected level) is
  not enough to satisfactorily explain the data. Then we tried some
  flavour-specific NP to address the puzzle, and found a much improved
  solution.

We take all the six modes, three each for $\Delta S=0$ and $\Delta S=1$, and perform an analysis based on the
SM only. We find that (i) the quality of the fit does not at all depend on whether the U-spin is exact or moderately 
broken, but (ii) it severely depends on how seriously we take the hierarchy among the polarisation fractions. If we
consider LPF dominance as suggested by theoretical models, 
the fit becomes abysmally poor. On the other hand, if we keep polarisation fractions as 
free parameters, the fit becomes excellent, but the hierarchy is no longer respected. 
The transverse polarisation amplitudes 
become comparable, or even bigger, than the LPFs, which is expected from the data. 

Given this conundrum, we entertain the idea that there might be some new physics that lowers the LPFs for the 
$\Delta S = 1$ modes. We take a bottom-up approach, and use a simplified effective theory where NP appears 
in $b\to s$ transitions, but not in $b\to d$ transitions. To make the analysis as simple as possible, we
assume exact U-spin symmetry among the SM parts of the amplitudes, as well as the hierarchy
$\left\vert A_-/A_0\right\vert = 0.3$ between the polarisation fractions as dictated by naive factorisation within SM. 
No such hierarchy has been assumed for the NP part, which is natural, as Lorentz structures other than the SM
one do not reproduce the hierarchy. 

We find a very significant improvement in the fit for almost all types of NP, even without considering a nonzero weak
phase in the NP Hamiltonian. The reason is obvious: the new 
Lorentz structures do not respect the hierarchy and hence can lower the LPFs. We have also checked that the 
fit results to other observables, namely, the CP-averaged branching ratios and CP asymmetries, are consistent
with the data. Finally, we make predictions for the LPF of the yet-to-be-observed mode $B_s\to K^{*+}K^{*-}$,
which has the power to differentiate among various NP options, as can be seen in Table~\ref{tabVIb}. We expect
our experimental colleagues to observe and measure this mode in near future.

\section*{Acknowledgement}
We thank Matthew D.\ Monk for pointing the latest LHCb result on $B_s\to\kst\kstbar$ to us.
DC and AK acknowledge the ANRF, Government of India, for support
through the projects CRG/2023/008234 and CRG/2023/000133 respectively.
DC also acknowledges the IoE, University of Delhi grant
IoE/2025-26/12/FRP.  SK acknowledges financial support through the
ANRF National Postdoctoral Fellowship (NPDF) with project grant no
PDF/2023/000410.

\appendix

\section{Form factors}\label{app}
The vector and axial-vector form factors for $B\to V_2$ are defined as
\begin{eqnarray}
    \left\langle V_2(k,\epsilon_2)|\bar{q}\gamma_{\mu}b|\bar{B}(p) \right\rangle &=& -i\varepsilon_{\mu\nu\alpha\beta}\epsilon^{*\nu}_2p^{\alpha}k^{\beta}\frac{2\mfv(q^2)}{m_B + m_{V_2}}, \nonumber\\
    \left\langle V_2(k,\epsilon_2)|\bar{q}\gamma_{\mu}\gamma_5 b|\bar{B}(p) \right\rangle &=&\epsilon^*_{2\mu} (m_B + m_{V_2})\mfa_1(q^2) - (p+k)_{\mu}(\epsilon^*_2.q)\frac{\mfa_2(q^2)}{m_B + m_{V_2}} \nonumber\\
    && -q_{\mu}(\epsilon^*_2.q)\frac{2m_{V_2}}{q^2}\left[\mfa_3(q^2)-\mfa_0(q^2)\right],
\end{eqnarray}
with \begin{equation}
   \mfa_3(q^2) = \frac{m_B+m_{V_2}}{2m_V}\mfa_1(q^2)-\frac{m_B-m_{V_2}}{2m_V}\mfa_2(q^2).
\end{equation}
The scalar matrix element is zero whereas the pseudoscalar matrix element is given by
\begin{equation}
    \left\langle V_2(k,\epsilon_2)|\bar{q}\gamma_{5}b|\bar{B}(p) \right\rangle = -(\epsilon^*_2.q)\frac{2m_{V_2}}{m_b+m_q}\mfa_0(q^2).
\end{equation}
The tensor contributions can be parametrised as
\begin{eqnarray}
    \left\langle V_2(k,\epsilon_2)|\bar{q}\sigma_{\mu\nu}b|\bar{B}(p) \right\rangle &=& \varepsilon_{\mu\nu\alpha\beta}\left[\epsilon^{*\alpha}_2(p+k)^{\alpha}g_+(q^2) + \epsilon^{*\alpha}_2q^{\beta}g_-(q^2) + (\epsilon^*_2.q)p^{\alpha}k^{\beta}g_0(q^2)\right], \nonumber\\
    \left\langle V_2(k,\epsilon_2)|\bar{q}\sigma_{\mu\nu}\gamma_5b|\bar{B}(p) \right\rangle &=& i\left[\{\epsilon^{*}_{2\mu}(p+k)_{\nu}-(p+k)_{\mu}\epsilon^*_{2\nu}\}g_+(q^2) \right.\nonumber\\
    && \left. + \{\epsilon^*_{2\mu}q_{\nu}-q_{\mu}\epsilon^*_{2\nu}\}g_-(q^2) + (\epsilon^*_2.q)\{p_{\mu}k_{\nu}-k_{\mu}p_{\nu}\}g_0(q^2) \right]
    \end{eqnarray}
where $g_{\pm,0}$ are related to commonly used $T_{1,2,3}$ form factors by following
\begin{eqnarray}
    g_{+}(q^2) &=& -T_1(q^2),\nonumber\\
    g_{-}(q^2) &=& \frac{m^2_B-m^2_{V_2}}{q^2}\left[T_1(q^2) - T_2(q^2)\right],\nonumber\\
    g_{0}(q^2) &=& \frac{2}{q^2}\left[T_1(q^2) - T_2(q^2) - \frac{q^2}{m^2_{B}-m^2_{V_2}}T_3(q^2)\right]
\end{eqnarray}
We use the LCSR prescription to evaluate the form factors. We use the expansion in the exact same form as described in Ref.~\cite{Bharucha:2015bzk}
\begin{equation}
    \mff_i(q^2) = \frac{1}{1-q^2/m^2_{R,i}}\sum^N_{n=0}\alpha^i_n\left[z(q^2)-z(0)\right]^n,
\end{equation}
where \begin{equation}
    z(q^2) = \frac{\sqrt{t_+-q^2}-\sqrt{t_+-t_0}}{\sqrt{t_+-q^2}+\sqrt{t_+-t_0}}, \quad {\rm and} \quad  t_0 =t_+\left(1-\sqrt{1-\frac{t_-}{t_+}}\right).
\end{equation}
Here $t_{\pm} =(m_B\pm m_{V_2})^2$, the masses of resonances $m_{R,i}$ and the expansion parameters $\alpha^i_n$ can be found in Ref.~\cite{Bharucha:2015bzk}. We calculate the $B\to \rho$ and $B \to K^{*0}$ form-factors at $q^2 = m^2_{K^0}$ and $q^2 = m^2_{\overline{K}^{*0}}$ respectively. 
The matrix elements from a vacuum to a vector meson are parametrised as follows 
\begin{eqnarray}
    \left\langle V_1(q,\epsilon_1)|\bar{q}'\gamma^{\mu}q|0 \right\rangle &=& f_{V_1} m_{V_1}\epsilon^{\mu}_1 \nonumber\\
    \left\langle V_1(q,\epsilon_1)|\bar{q}'\sigma^{\mu\nu}q|0 \right\rangle &=& if^T_{V_1}(q^{\mu}\epsilon_1^{\nu}-\epsilon^{\mu}_1q^{\nu}),
\end{eqnarray}
where $f_{V_1}$ and $f^T_{V_1}$ are the decay constants of $V_1$ meson. The matrix elements for the remaining Lorentz structures are zero.


\end{document}